\documentclass[%
 aip,
 jmp,%
 amsmath,amssymb,
 reprint,%
]{revtex4-1}

\usepackage{graphicx}
\usepackage{dcolumn}
\usepackage{bm}

\begin{document}

\preprint{AIP/123-QED}

\title{High quality factor, fully switchable THz superconducting metasurface}

\author{G. Scalari}
 \altaffiliation[]{scalari@phys.ethz.ch}
\author{C. Maissen}%
\affiliation{ 
Institute of Quantum Electronics, Eidgen\"ossische  Technische Hochschule Z\"urich, Switzerland}
\author{S. Cibella, R. Leoni}
\affiliation{Istituto di Fotonica e Nanotecnologie (IFN), CNR, via Cineto Romano 42, 00156 Rome, Italy}
\author{J. Faist}
\affiliation{ 
Institute of Quantum Electronics, Eidgen\"ossische  Technische Hochschule Z\"urich, Switzerland}

\date{\today}

\begin{abstract}
We present a complementary THz metasurface realised with Niobium thin film which displays a quality factor Q=54 and a fully switchable behaviour as a function of the temperature. 
The switching behaviour and the high quality factor are due to a careful design of the metasurface aimed at maximising the ohmic losses when the Nb is above the critical temperature and minimising the radiative coupling. The superconductor allows the operation of the cavity with an high Q and  inductive elements with an high aspect ratio. Comparison with three dimensional finite element simulations highlights the crucial role of the inductive elements and of the kinetic inductance of the Cooper pairs  in achieving  the high quality factor and the high field enhancement. 
\end{abstract}

\maketitle

Metamaterials  \cite{Pendry:IEEE:99,SchurigSci:06:977,Chen:Nat:2006} proved in the last 15 years to be an extremely flexible and useful concept \cite{Engheta:Sci:07} that has been implemented  in many different contexts, from fundamental research \cite{papasimakis:PRL:08} to applications \cite{Belkin:Nat:14}. The possibility to engineer the electromagnetic response of surfaces  and materials \cite{CaiShalaev:NATPHOT:07} with 2D \cite{Halasreview:NATMAT:10} and 3D structures \cite{GanselWegener:Sci:09,Paulillo:OE:14,FanAveritt:OE:11} with constitutive elements much smaller than the wavelength has opened new and exciting  possibilities, from integrated optics \cite{Yu:Sci:11}  to solid state physics \cite{LiuAveritt:Nat:12} and strong light-matter coupling\cite{scalari:science:2012}, just to name a few.

 Recently, we reported on experiments on ultrastrong light-matter coupling at THz frequencies with superconducting Niobium (Nb) metasurfaces \cite{Scalari:NJP:14}. In order to address theoretical predictions on peculiar cavity QED effects which arise in such ultrastrong coupling regime \cite{Ciuti:PRB:05:115303-1,Ridolfo:PRL:13,Ridolfo:PRL:11,Stassi:PRL:13}, it is necessary to be able to perform non-adiabatic experiments at THz frequencies. Superconducting metamaterials have been investigated in the THz \cite{Ricci:APL:05,Jin:OE:10,Zhang:OE:12,Chen:PRL:10,Anlage:JOP:11} and superconductors (SC) indeed display transition times of the order of the ps  \cite{Singh:Nanophot:12,matsunaga:PRL:12}  when illuminated with high intensity, ultrafast optical pulses.
The objective of our study is the realisation of a metasurface operating in the THz with high quality  factor and displaying a pronounced switching behaviour. We would like to achieve a metasurface that displays a two-state behaviour: a narrow band, high transmission in the range 200-400 GHz (well below the Nb gap frequency of 730 GHz) that can be commuted to a broadband, low transmission.  At the same time, since we are interested in cavity QED at THz frequencies \cite{scalari:science:2012}, we aim at realising a meta-atom which provides a very small mode volume, in order to enhance the vacuum field fluctuations that scale as $E_{vac} \sim \sqrt{\frac{\hbar\omega}{\epsilon V_{cav}}}$ where $V_{cav}$  is the cavity mode volume, $\omega$ is the photon angular frequency and  $\epsilon=n_{eff}^2$ where n$_{eff}$ is the effective refractive index  \cite{devoret2007}.  The same high ratio $\frac{\omega}{V_{cav}}$ will lead to an high  field enhancement factor.

Switchable metasurfaces have been presented in literature, by using composite metamaterials structures \cite{HTChen:NatPhot:08} or by employing BCS or high T$_c$ superconductors \cite{Jin:OE:10,Zhang:OE:12,Chen:PRL:10,Singh:APL:13}. Q factors of these structures were in the range of  5-15, and in some cases \cite{Chen:PRL:10,Singh:APL:13} the resonance was efficiently quenched for temperature values above T$_c$.
In our previous structures \cite{Scalari:NJP:14}, we obtained a switching ratio $\frac{Q_{T<T_c}}{Q_{T>T_c}}=\frac{8.2}{2.46}=3.3$ between  3 K and 10 K  with a non-optimised structure.

\begin{figure}[h]
\begin{center}
   \includegraphics[width=90mm]{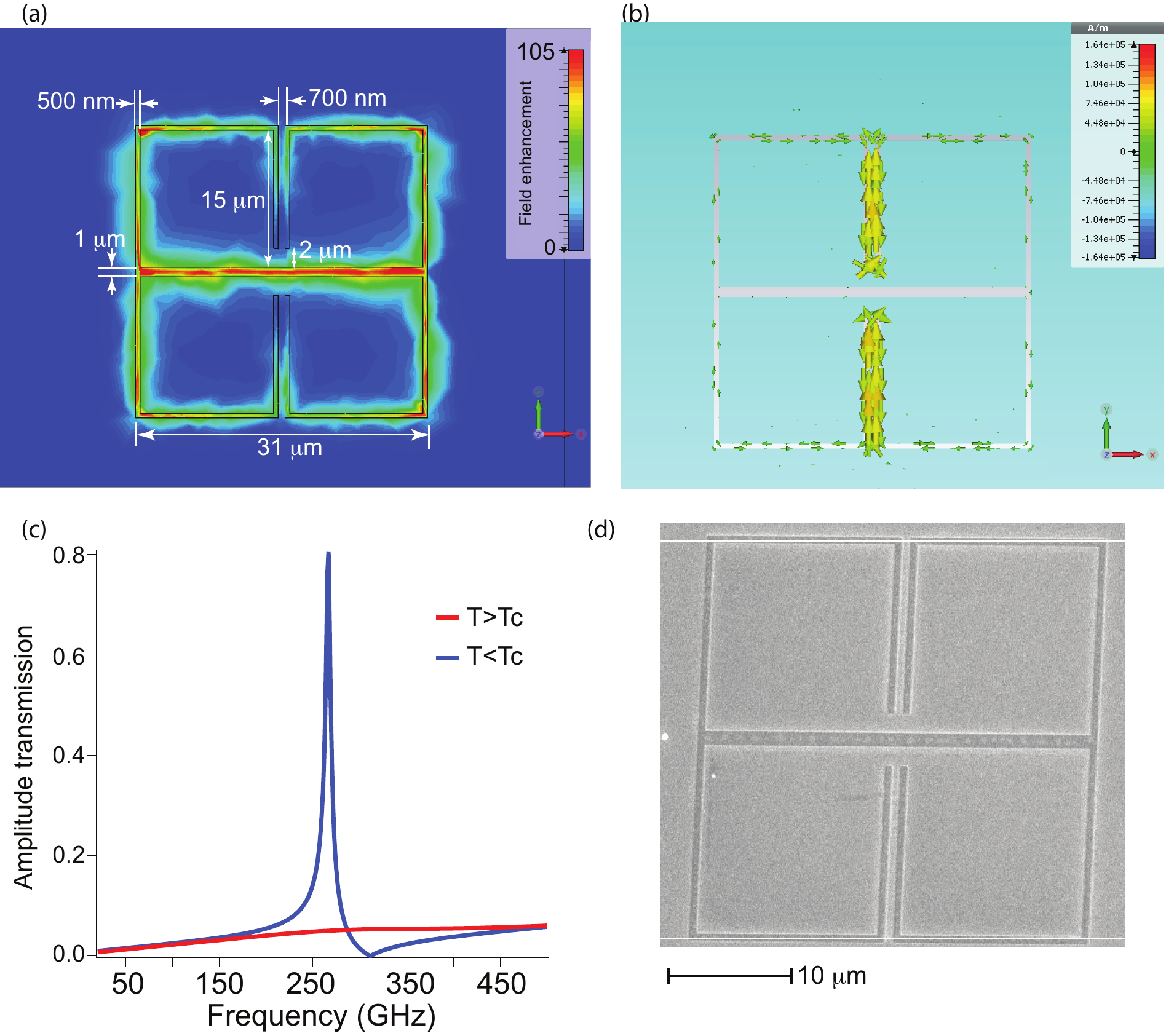}
    \caption{(a): Total electric field intensity ratioed to the incident field in the simulation for the resonator in the superconducting state simulated with $T<T_c$ for the resonant frequency $\nu= 266 $ GHz.  (b): Simulated surface current at the LC resonance $\nu= 266 $ GHz. (c): Simulated amplitude transmission for an array of resonators above and below the critical temperature. The square unit cell has a side of 60 $\mu$m.  (d): SEM picture of the fabricated sample with 150 nm thick Nb metasurface on a Semi-Insulating GaAs substrate.  } 
\label{resscheme}
   \end{center}
\end{figure}

In the present study we used Nb as a superconducting material and we designed the cavity to be resonant around 250 GHz.
  In order to maximise the switching effect due to the presence of the superconducting state, we adopted the following design strategy: we increased the radiative Q factor in order to have a structure whose resonance line width is not radiatively limited but loss limited. Then, as a second step, we engineered the  inductive elements using a very high aspect  ratio $\frac{l}{A}$ (l length and A cross section)in order to maximise the losses when the structure is in the normal state.  At the same time these elements will contribute negligibly to the losses when the Nb is superconducting. The resonator design is reported in Fig.\ref{resscheme}  together with 3D simulations of the electric field distribution, the surface currents, the simulated amplitude transmission and an SEM picture of the fabricated metasurface. The radiative quality factor can be engineered by acting on the capacitor gap dimension  in order to reduce the efficiency of the dipolar coupling  as well as the inter-meta-atom spacing \cite{Singh:APL:10:241108}. When designing the structure we then identified the regions of the resonator were the highest current density is flowing (see Fig.\ref{resscheme}(b)) , which will give rise to the higher ohmic losses. We then engineered such regions in order to maximize the effect of the superconducting transition. By employing very narrow (700 nm) and long (13 $\mu$m) inductive elements we obtain a very high resistance when the superconductor is above T$_c$ (for our film $T_c=8.7$ K \cite{Scalari:NJP:14}). On the contrary, when the Nb is operated below T$_c$, such regions present an higher inductance due to the kinetic inductance of the Cooper pairs and their loss is reduced by the very low, close to zero, value of the real part of the surface impedance.   
  
 In order to simulate the structure we employed a surface impedance model as reported in Refs \cite{Chen:PRL:10,Scalari:NJP:14}. The surface impedance values for the Nb have been measured in our previous study \cite{Scalari:NJP:14}.  The detailed geometry of the meta-atom is reported in Fig.\ref{resscheme}(a) and in the caption. From the simulations of Fig.\ref{resscheme}(a) we can see that we obtain an high field enhancement factor of about 100 at resonance and from the predicted amplitude transmission curves (Fig.\ref{resscheme}(c)) we can see that the designed metasurface provides the searched switchable behaviour. 
The metasurface is fabricated with Electron-Beam lithography and Reactive Ion Etching  on a 100 nm thick Nb film following the same procedure reported in \cite{Scalari:NJP:14}: an SEM micrograph of the fabricated sample is visible in Fig.\ref{resscheme}(d).
 
\begin{figure}[h]
\begin{center}
   \includegraphics[width=85mm]{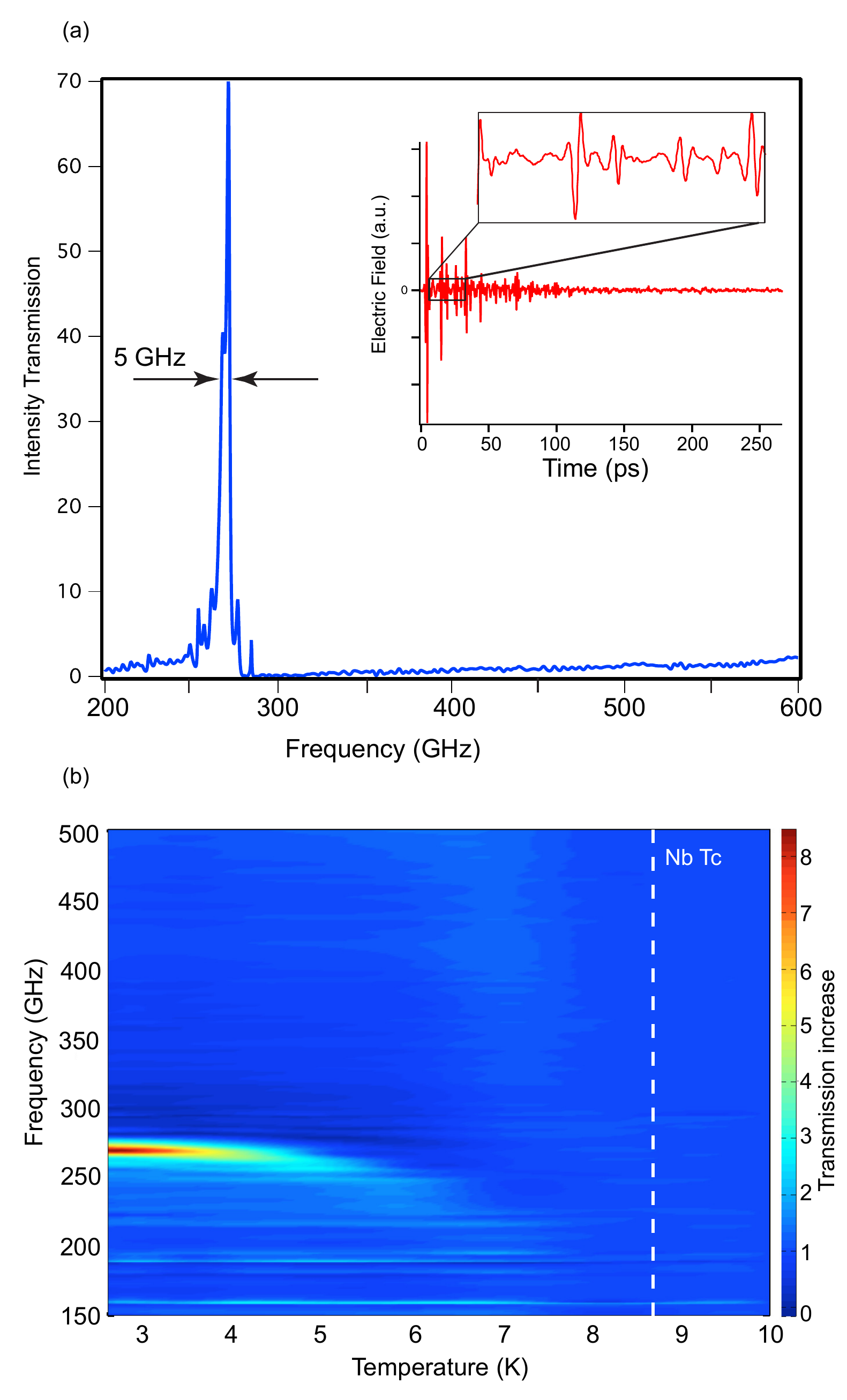}
    \caption{(a): Metasurface intensity transmission at T=2.6 K normalised to the transmission value at T=10 K for a long scan of 40 mm.  Inset: time trace of the scan (b): color plot of the metasurface amplitude  transmission as a function of the temperature, normalised to the transmission at 10 K.  } 
\label{measVsT}
   \end{center}
\end{figure}

\begin{figure}[h]
\begin{center}
   \includegraphics[width=80mm]{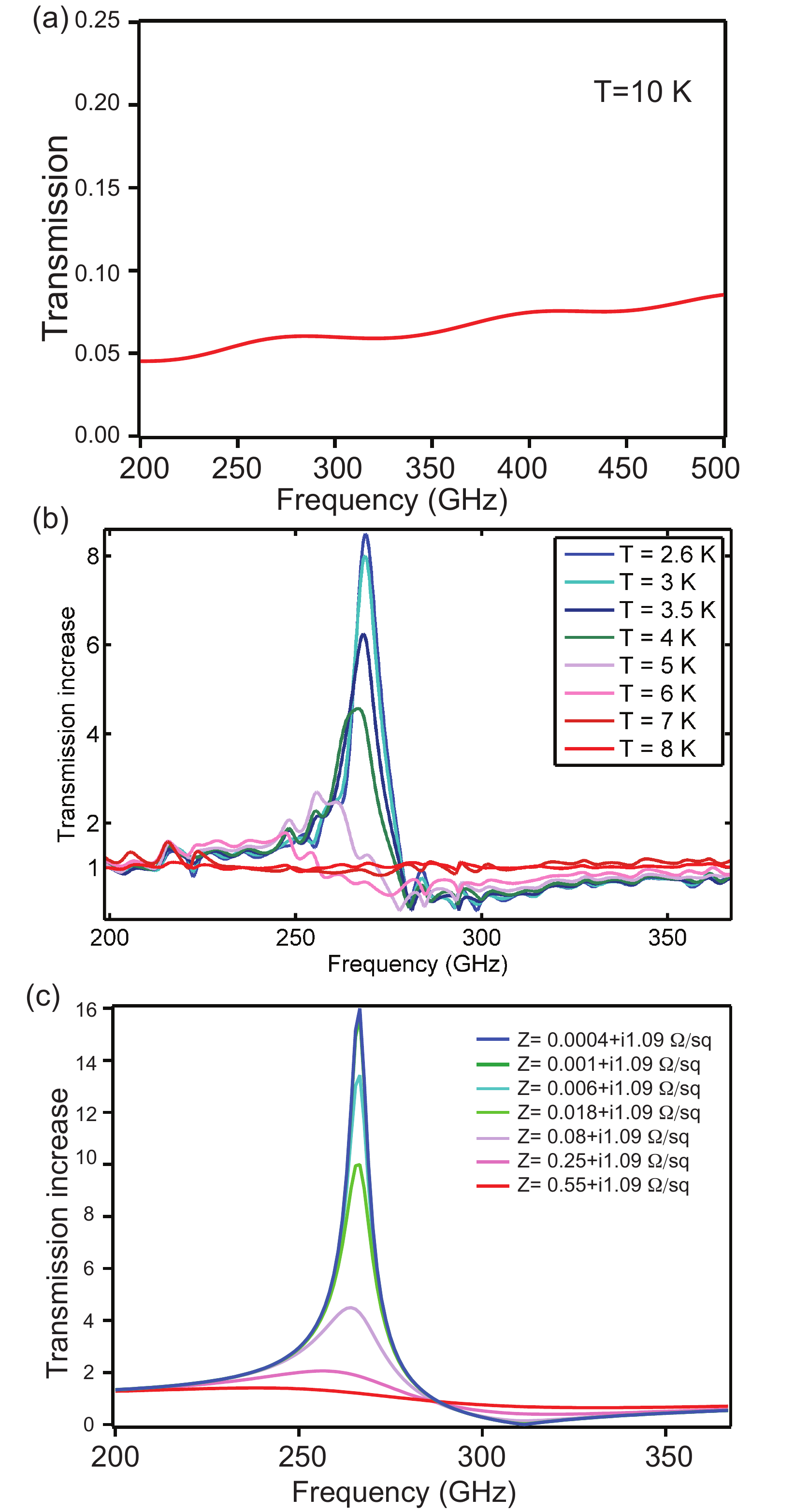}
    \caption{(a): Metasurface transmission at 10 K referenced to empty sample holder.  (b): metasurface amplitude transmission as a function of the temperature, normalised to the 10 K transmission. In this case the scan is shorter with respect to Fig.\ref{measVsT}(a)  and the linewidth is limited by the scan resolution.  (c): simulation of the metasurface amplitude transmission normalised to the 10 K simulated transmission as function of different values of surface impedance.  } 
\label{measVsTsimul}
   \end{center}
\end{figure}

 Measurements are performed with a THz-TDS system (described in Ref\cite{scalari:science:2012} ) coupled to a bath cryostat where we can vary the temperature of the sample between 2.6 K and 300 K.   In Fig. \ref{measVsT}(a) we report the transmission spectrum for the metasurface at the lowest temperature T=2.6 K. The normalisation of the the transmission is made using the high temperature (10 K) transmission of the metasurface, when the Nb is fully in the normal state an shows no resonance (see Fig.\ref{measVsTsimul}(a)). This allows to perform a high resolution measurement reducing the interference fringes due to multiple reflections of the THz pulses inside the sample. The measured quality factor at T=2.6 K is Q=54, which compares well with results obtained  on other superconductor metamaterials \cite{Jin:OE:10,Zhang:OE:12,Chen:PRL:10} and is comparable with the best results on metamaterials employing interference phenomena  \cite{Cao:OL:12}. The structure displays a very narrow transmission resonance peaked at 269 GHz for the lowest temperature. When the sample is heated the transmission resonance broadens and redshifts and finally disappears for temperatures above 8 K (see Fig.\ref{measVsTsimul}(a)). This behaviour can be appreciated by looking at the colour plot of Fig. \ref{measVsT}(b).
  The observed behaviour as a function of the temperature  is qualitatively well reproduced by the simulations  reported in Fig.\ref{measVsTsimul}(c).  

As it clear from simulations and measurements, the design of the metasurface was optimised as well with respect to the symmetry and the shape of the resonance. In other high-Q design approaches based on inter-meta atoms coupling, interference phenomena and Fano-like coupling\cite{Cao:OL:12,Halasreview:NATMAT:10,papasimakis:PRL:08,Tassin:PRL:09,Gu:Natcomm:12} the line shape of the high Q mode is,by construction, frequently complex and spectrally very close to other broader resonances. If employed in cavity QED experiments, this kind of line shape could result in complex spectra which would be not so easy to interpret.
 
\begin{figure}[h]
\begin{center}
   \includegraphics[width=90mm]{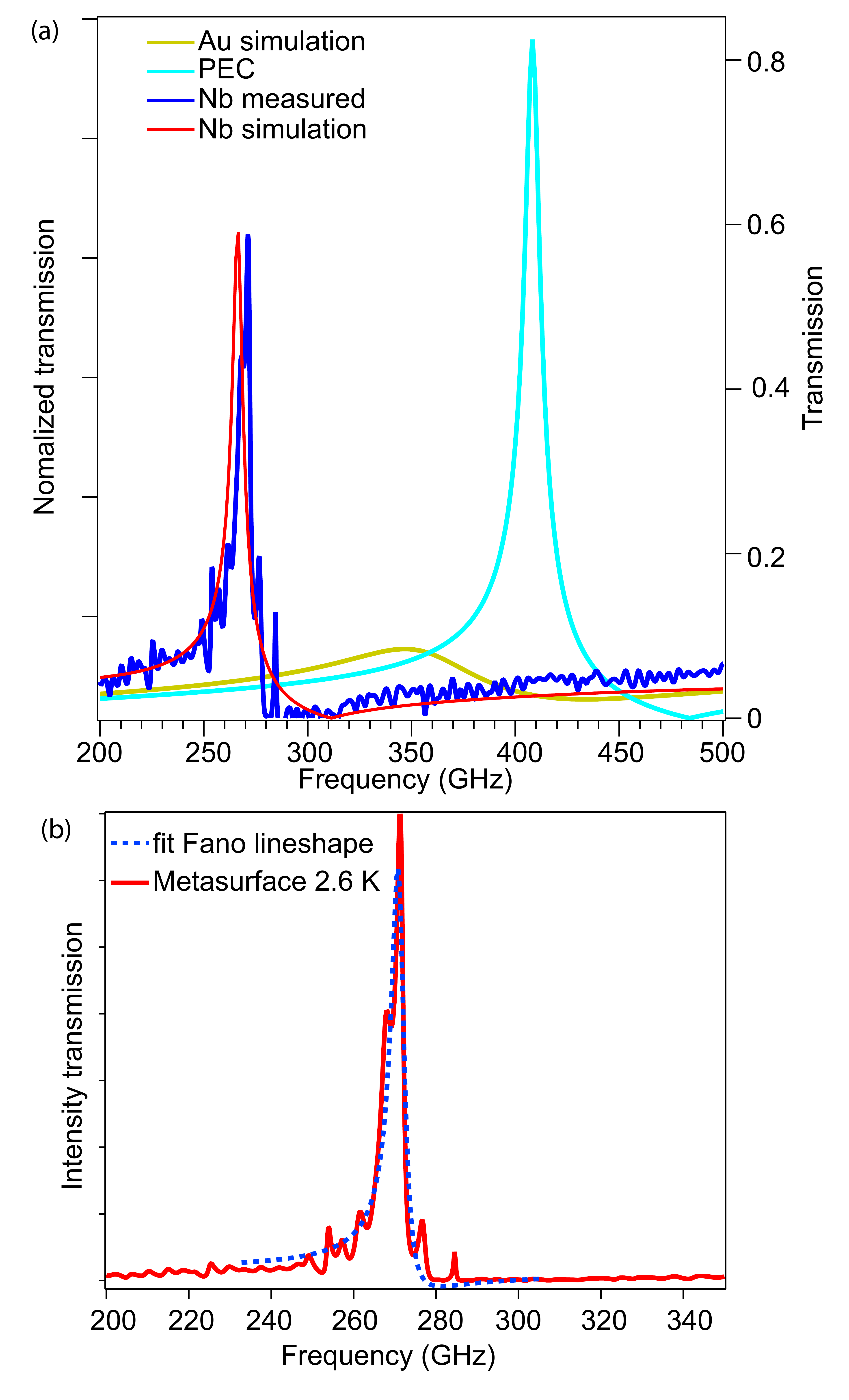}
    \caption{(a): Experimental transmission at 2.6 K normalised to transmission at 10 K for the Nb metasurface (blue line, left y-axis) together with simulated curve for the same quantity (red line, left y-axis, the surface impedance for the Nb is $Z_{Nb}=0.0004+i1.09)$ $\Omega/Sq$).    Simulated transmission for Perfect Electric Conductor (light blue line, right y-axis) and Gold (yellow line, right y-axis).  (b): Intensity transmission for the metasurface at 2.6 K  together with a fit employing the Fano line shape. } 
\label{simulcompar}
   \end{center}
\end{figure}

 We now want to analyze the behaviour of the structure using finite element simulations performed with CST microwave studio. We performed simulations  employing Nb, Au and Perfect Electric Conductor (PEC) as materials for the metasurface. The Gold has been also simulated with the surface impedance method in order to be fully comparable with the Nb. The results of such simulations are reported in Fig. \ref{simulcompar}(a).  The PEC simulation allows us to disentangle the roles of the radiative quality factor Q$_{rad}$ and of the loss quality factor Q$_{loss}$. Since PEC does not present any material loss the quality factor of the structure in this case represents the radiative coupling and is equal to Q$_{rad}=62$. 
 
 The high radiative quality factor in this kind of electronic split ring resonator is the result of our optimised design for a low loss conductor. We chose the symmetric, electronic-like \cite{Schurig:APL:06} split ring resonator approach in order to control the radiative coupling mainly with the capacitor gap and its width. Due to its symmetry, the coupling to the resonator through magnetic field is negligible and the coupling happens at first order through the capacitor. By adopting a narrow (1 $\mu$m gap ) capacitor we can increase the radiative quality factor and  target a fairly low frequency using long inductors. The use of long inductive elements would be penalising because of the high ohmic loss, but in our case we employ a superconductor so we keep these losses at the minimum.  
The relation which expresses the total quality factor for a resonator  as a function of material loss and radiation loss is:

\begin{eqnarray}
\frac{1}{Q_{tot}}=\frac{1}{Q_{rad}}+\frac{1}{Q_{loss}}
\end{eqnarray}
 
We can then use the simulated value of the quality factor for the PEC to deduce the loss quality factor for the measured Nb and the simulated  Au. All the results are reported in Tab.\ref{table: Qfactor} I. The resonator we propose, if fabricated with Gold, presents a very low loss quality factor $Q_{loss}^{Au}=5.4$ due to the narrow and long  wires that result highly dissipative even when fabricated with a good conductor.  On the contrary, the Nb resonator displays a high loss quality factor  $Q_{loss}^{Nb}=418$ when in the superconducting state due to the low value of the real part of the surface impedance and the high value of the reactance. 

 It is important to highlight the narrow temperature range ($\Delta T=4$ K) over which the switching of the overall Q factor of the resonator occurs, compared to other results reported in literature for high T$_c$ superconductors \cite{Chen:PRL:10, Singh:APL:13}.  
 
 It is interesting to note also the important contribution of the kinetic inductance of the superconducting phase to determine the correct frequency of the resonator. If we examine again Table I we see that for a perfect electric conductor with no kinetic inductance the resonant frequency is $\nu_{PEC}=408 $ GHz which is 1.5 times of what obtained with the Nb. This highlights the role of the term $L_{k}$ in the expression for the resonance frequency of the superconducting metasurface below T$_c$ expressed as:  
 
 \begin{equation}\label{freqLCsuper}
\nu_{res}=\frac{1}{2\pi}\sqrt{\frac{1}{(L_g+L_k)C}}
\end{equation}
where $L_g$ is the geometric inductance and C the capacitance of the LC circuit associated to the meta-atom.
The kinetic inductance for a superconducting wire of length $l$ and cross sectional area $A$ can be written as :
 \begin{equation}\label{kineticL}
L_k=\frac{m}{2 n_s e^2}\frac{l}{A}
\end{equation}
where $n_s$ is the Cooper pair density and m the electron mass.
 We already pointed out that we chose inductances with a very high aspect ratio $\frac{l}{A}=\frac{13 \times 10^{-6}}{100 \times 10^{-9} \times 700 \times 10^{-9}} \simeq 1.8\times 10^8$ in order to maximise the resistance when $T>T_c$. The same aspect ratio enhances the kinetic inductance role in the determination of the resonator frequency.  
 
\begin{table}[h]
\label{table: Qfactor}
  \centering	
\begin{tabular}{|c|c|c|c|c|}
  \hline
  &  Nb Theo & Nb exp & Au theo &  PEC   \\
  \hline
    \hline
    $ \nu$ (GHz) & 266 &  269 & 346 & 408  \\
        \hline
   \hline
     Q$_{tot}$ & 53 & 54 & 5 & 62  \\
      \hline
 Q$_{rad}$ &62  & 62  &62  & 62  \\
       \hline
 Q$_{loss}$ &  405 & 418   &5.4  & $\infty$  \\
  \hline
\end{tabular}
\caption{Simulated and experimental quality factor and resonant frequencies for the Nb sample below T$_c$ and simulated values for Au and Perfect electric conductor resonator with the same geometry of the Nb. The surface impedance value used for Au is $Z_{Au}=0.135 (1+i)$ $\Omega/Sq$ }\label{allsamplestable}
\end{table}

In the case of Gold and PEC the kinetic term is very small or absent \cite{Singh:APL:13} and the resulting resonator frequency is much higher. 
This fact has a strong  impact when considering the field enhancement factor and the vacuum field fluctuation for a cavity at a given frequency. The lowering of the ohmic loss introduced by the PEC condition allows to have a high Q$_{loss}$ but is the presence of the high kinetic inductance term of the Cooper pairs which leads to a very favourable frequency-to-volume ratio.

The  split-ring resonator cavity, whose dimensions are already strongly sub wavelength, is now yielding a much lower frequency because of the kinetic inductance contribution which lowers the frequency according to Eq. \ref{freqLCsuper} and as visible from the simulation values  in Table I. In order to obtain such a low frequency with a normal conductor, we should have rescaled the resonator with bigger dimensions, then lowering the ratio $\frac{\omega}{V_{cav}}$: the kinetic inductance is instrumental in obtaining a more compact resonator \cite{Anlage:JOP:11}.  From our 3D simulations we deduce a field enhancement of $E^{enhance}_{Nb}=105$ at $\nu=269$ GHz for the superconducting resonator (see Fig.\ref{resscheme}(a)). 
%
%

%
If one inspects the line shape of the metasurface resonance it is evident a slight asymmetry with a steeper slope on the blue side. In Figure \ref{simulcompar}(b) we plot again the intensity transmission for the metasurface at 2.6 K together with the fit employing a Fano intensity line shape \cite{Halasreview:NATMAT:10}. The good agreement of the fit with the data (Fano coefficient f=0.19, $\gamma=2$ GHz) reveals the underlying Fano coupling between the low Q, highly spatially delocalised weak transmission of the array and the high Q, LC resonance of our complementary split-ring cavities. 

In conclusion, we demonstrated a new design for a complementary superconducting cavity which presents a high switching ratio between the superconducting state, where the quality factor of the resonance is as high as 54, and the normal state where the resonance is absent. The role of the kinetic inductance and of the high aspect ratio inductive element is elucidated with measurements and finite element simulations. This kind of cavity can be applied in cavity quantum electrodynamics at THz frequencies and as a narrow band switchable filter.

We acknowledge financial support from the Swiss National Science Foundation (SNF) through the National Centre of Competence in Research Quantum Science and Technology and through the SNF grant n. 129823. We also acknowledge financial support from the ERC grant MUSIC.  We would like to thank J. Keller for help with measurements and F. Valmorra for the careful reading of the manuscript.

%

\end{document}